\documentclass{article}
\usepackage[utf8]{inputenc}
\usepackage{authblk}
\usepackage{hyperref}
\usepackage{bm}
\usepackage{amsmath}
\usepackage{appendix}

\title {Einstein’s Hidden Scaffolding, with a Glance at Poincaré}

\author{Galina Weinstein\thanks{The Department of Philosophy, University of Haifa.}}

\begin{document}

\maketitle

\begin{abstract}
This paper reconstructs the derivations underlying the kinematical part of Einstein's 1905 special relativity paper, emphasizing their operational clarity and minimalist use of mathematics. Einstein employed modest tools—algebraic manipulations, Taylor expansions, partial differentials, and functional arguments—yet his method was guided by principles of linearity, symmetry, and invariance rather than the elaborate frameworks of electron theory. The published text in \emph{Annalen der Physik} concealed much of the algebraic scaffolding, presenting instead a streamlined sequence of essential equations. Far from reflecting a lack of sophistication, this economy of means was a deliberate rhetorical and philosophical choice: to demonstrate that relativity arises from two simple postulates and basic operational definitions, not from the complexities of electron theory. The reconstruction highlights how Einstein’s strategy subordinated mathematics to principle, advancing a new mode of reasoning in which physical insight, rather than computational elaboration, held decisive authority. In this respect, I show that Einstein’s presentation diverges sharply from Poincaré’s. 
\vspace{2mm}

\noindent \textbf{This paper is in memory of John Stachel, whose life’s work was devoted to illuminating Einstein’s special and general relativity.}
\end{abstract}

\section{Introduction}

This paper reconstructs the derivations underlying the kinematical part of Einstein’s 1905 relativity paper, making explicit the operational program and the modest mathematics that support it. Einstein works with lean tools, algebraic manipulations, Taylor expansions, partial differentials, and functional arguments, but the real drivers are principles: linearity, homogeneity, isotropy, and invariance. Rather than relying on the elaborate machinery of electron theory, he shows how the relativity principle and the light postulate, together with clear operational definitions, are sufficient.

The \emph{Annalen der Physik} text largely suppresses the algebraic scaffolding, presenting a sequence of essential equations pared to their conceptual core. This economy is not a deficit of technique but a deliberate methodological choice: the mathematics is kept subordinate to principle, so that the theory’s force appears to spring from two postulates and straightforward measurement procedures with rods and clocks. In this respect, I show that Einstein’s presentation diverges sharply from Poincaré’s. 

\section{Simultaneity and Synchronization}

\subsection{Simultaneity in One Inertial Frame} \label{RS}

Consider clocks at $A$ and $B$ at rest in the same inertial system. If $A$ emits a light signal at $t_A$, $B$ reflects it at $t_B$, and $A$ receives it at $t'_A$, then $A$’s and $B$’s clocks are synchronized if \cite{Einstein05}: 

\begin{equation} \label{eq114}
\boxed{t_B - t_A \;=\; t'_A - t_B.}   \qquad \text{A definition of synchronization.}
\end{equation}
This equation says: we shall call clocks synchronized if the signal travel time out equals the signal travel time back.
Notice that logically speaking, Einstein \emph{did not} derive this relation. He posited it as the operational meaning of simultaneity between distant clocks in a single inertial frame.

\noindent Equation \eqref{eq114} can be rearranged:

\begin{equation}\label{eq1}
2t_B \; = \; t_A + t'_A \rightarrow \quad  t_B = \frac{t_A + t'_A}{2}. 
\end{equation}

Einstein was attempting to define distant simultaneity within a single inertial frame of reference. 
Ether-drift experiments only established the two-way (round-trip) speed of light.%
\footnote{Ether-drift experiments intended to test whether Earth’s motion relative to the ether produces measurable anisotropy in light propagation. But these only established the two-way (round-trip) speed of light. They showed no detectable directional dependence. In 1849, Hippolyte Fizeau conducted an experiment that determined the speed of light using a toothed-wheel method. It measured the two-way speed of light. It was not designed to test the ether theories, but like all round-trip methods, it could not, in principle, access the one-way speed. Thus, both Fizeau’s 1849 determination of the velocity of light and the ether-drift experiments established only the two-way speed of light. Neither provided direct access to the one-way speed, which remains underdetermined without a convention for synchronizing clocks at a distance.}
Without further assumptions, it is logically consistent to imagine anisotropic one-way speeds. Light might go from $A$ to $B$ at speed $c_1$ and return from $B$ to $A$ at speed $c_2$, as long as the harmonic mean satisfies $c$. So the round-trip ether drift experiments would not rule that out, and the one-way speeds remained underdetermined by experiment. 
No direct experiment can measure the one-way speed of light without already having synchronized clocks at spatially separated points. Einstein’s definition of simultaneity \eqref{eq114} fixes the synchronization rule by stipulation: the travel time out equals the travel time back. That is equivalent to assuming \emph{isotropy} (no preferred direction): within the chosen inertial frame, the light travel time is stipulated to be equal in both directions: 
\begin{equation*}
c_{\text{one-way}}, \, A \rightarrow B  =c_{\text{one-way}}, \, B \rightarrow A = c    
\end{equation*}
This removes the ambiguity: the one-way speed is no longer an open empirical question but a definition. It also brings operational clarity because simultaneity becomes measurable, not intuitive. Finally, this move allows the Lorentz transformations to be derived from the definition \eqref{eq114}, the relativity principle, and the light postulate.

Einstein lays down logical properties that his definition of clock synchronization \eqref{eq114} must satisfy if it is to make sense across a network of clocks \cite{Einstein05}:

1) \emph{Symmetry} (sometimes called reciprocity): If the clock at $B$ synchronizes with the clock at $A$, the clock at $A$ synchronizes with the clock at $B$. 

2) \emph{Transitivity}: the synchronization relation can be extended to more than two clocks consistently so that one can build a network of synchronized clocks. If $A$ is synchronized with $B$ and $C$, then $B$ is synchronized with $C$.

\subsection{The Two-Way Speed of Light is a Constant \texorpdfstring{$c$}{c}}

Next, Einstein connects the definition \eqref{eq114} with the empirical result that the round-trip light speed is $c$, and postulates (in agreement with two-way measurements) that the velocity of light in vacuum is the universal constant $c$ \cite{Einstein05}.  

\noindent The outward trip $A \to B$ (the light takes the time going from $A$ to $B$) is:
\begin{equation} \label{eq128}
\Delta t_{AB} = t_B - t_A.    
\end{equation}
Now we substitute the definition of $t_B$ equation \eqref{eq1} into equation \eqref{eq128} to get:
\begin{equation}
\Delta t_{AB} = \frac{t_A + t'_A}{2} - t_A = \frac{t'_A - t_A}{2}.
\end{equation}
The return trip $B \to A$ is $\Delta t_{BA} = t'_A - t_B$.
Now again we substitute $t_B$ \eqref{eq1}:
\begin{equation}
\Delta t_{BA} = t'_A - \frac{t_A + t'_A}{2} = \frac{t'_A - t_A}{2}.
\end{equation}
The result is that both halves are equal: 

\begin{equation} \label{eq130}
\Delta t_{AB} = \Delta t_{BA} = \frac{t'_A - t_A}{2}.
\end{equation} 
Both one-way times are equal, which allows us to express the speed either as $\frac{AB}{\Delta t}$ (one-way) or as $\frac{2AB}{t'_A - t_A}$ (round-trip), and they give the same result. $AB$ is the distance between the two points $A$ and $B$ in the reference system (the length of the straight line segment). 
Thus, the round-trip expression, the average speed applied to the round-trip path of the light signal, becomes \cite{Einstein05}: 

\begin{equation}\label{eq131}
v_{\text{avg}}=\boxed{\frac{2\,\overline{AB}}{t'_A - t_A}}=c.
\end{equation}
which matches the empirical constant. Einstein wrote $\overline{AB}$ to clarify that he is referring to the line segment from $A$ to $B$. Thus, the average velocity is the “distance there and back” (twice the segment $AB$) divided by the elapsed time. 

\noindent The operational outcome is that, in agreement with experience, the two-way round-trip speed of light is the universal constant $c$. $\rightarrow$ The postulate: Any ray of light moves within an inertial system with the velocity $c$ \cite{Einstein05}.

\subsection{Longitude, Latitude, and Attitude}

In "The Measurement of Time" (1898) and again in 1900, Henri Poincaré provided the first explicit philosophical and operational account of distant simultaneity. He described how to synchronize spatially separated clocks by exchanging light (or telegraph) signals, linking this procedure to Hendrik Antoon Lorentz’s \emph{local time}. The core reasoning was that observers in uniform motion, unaware of their translational velocity, would assume light to propagate isotropically \cite{Poi98, Poi00}. 

In 1898, Poincaré analyzed telegraphic longitude determinations between Paris and Berlin \cite{Poi98}. Although he did not explicitly write down or derive a midpoint formula \eqref{M}, the underlying logic is clearly present.%
\footnote{The procedure can be summarized as follows:
\begin{enumerate}
\item Paris sends a signal at $t_1$ (Paris time); Berlin receives it at $t_2$ (Berlin time) and replies at $t_3$; Paris receives the return signal at $t_4$.
\item Assuming isotropic transmission, the one-way travel time is:
\begin{equation} \label{M}
\frac{(t_4 - t_1) - (t_3 - t_2)}{2}.
\end{equation}
\item The corrected longitude difference is then obtained by subtracting the appropriately adjusted Berlin time from Paris time.
\end{enumerate} Poincaré never wrote down the midpoint formula in the neat algebraic form I have given in \eqref{M}. That is a modern reconstruction of the logic implicit in his 1898 discussion of telegraphic longitude determinations.}
He describes a procedure of exchanging signals, neglecting (or, if necessary, correcting for) the transmission delay, and treating the times as equal. In this sense, the principle of the midpoint rule is already embedded in his account, though presented as a practical convention of astronomy rather than as a universal definition of simultaneity.

In 1900, Poincaré described how observers at different points might synchronize their clocks by exchanging light signals. He noted that such observers, unaware of the Earth’s translational motion, would assume that signals propagate equally fast in both directions. This assumption amounts to treating the one–way speed of light as \emph{isotropic}—not based on experiment, but as a postulate built into the synchronization convention. If light propagation were truly anisotropic, correcting for transmission delays would require knowing the direction-dependent speeds. Instead, Poincaré's observers assume equality, thereby defining Lorentz's local time \eqref{eq14}. In this framework, the ether rest-frame observer employs the “true” time $t$; for them, the one–way speed of light is anisotropic in moving systems, though the two–way (round–trip) speed remains $c$. The moving observer, by contrast, presumes isotropy and synchronizes clocks accordingly, so that their clocks measure not the true time $t$, but the local time $t'$ \cite{Poi00}.  

So, Poincaré proposed that observers at different locations—for instance, $A$ and $B$—synchronize clocks as follows \cite{Poi00}:  
\begin{enumerate}
    \item $A$ sends a light signal to $B$.  
    \item Upon receipt, $B$ immediately returns a signal to $A$.  
    \item $A$ records emission and reception times, assuming equal transit times each way.  
\end{enumerate}
Under this assumption, the clocks are adjusted to agree. In reality, if the system moves through the ether, the forward and return times differ. Still, the observers’ ignorance of this motion leads them to adopt an \emph{apparent} local time $t'$:
\begin{equation} \label{eq14}
t' = t - \frac{x v}{c^2},
\end{equation}
differing from the “true” time $t$, with the offset depending on their velocity relative to the ether. 

Poincaré does not explicitly connect the 1898 midpoint convention with the 1900 local time construction. 

\subsection{Making Poincaré Appear “Almost Einstein”}

Einstein’s 1905 analysis unfolds in three distinct steps: 

First, he begins with an \emph{empirical input}: ether-drift experiments had already established that the two-way velocity of light is constant, independent of the source’s motion. 

Second, he introduces a \emph{definition}, stipulating that the one-way velocity of light be treated as isotropic by a convention of synchronization [equation \eqref{eq114}]. 

Third, he offers a \emph{synthesis}: with this stipulation, the measured round-trip velocity coincides with the universal constant $c$. 

In so doing, Einstein elevates $c$ from a mere optical parameter to a postulate of kinematics. Crucially, synchronization itself becomes a universal operational definition of simultaneity—valid in every inertial frame. What had been an astronomical convenience becomes the very cornerstone of the new kinematics.

According to Arthur Miller’s book, \emph{Albert Einstein’s Special Theory of Relativity: Emergence (1905) and Early Interpretation, 1905–1911} \cite{Miller}, Poincaré had already outlined elements of the problem—he recognized the fragility of simultaneity, identified logical difficulties, and suggested partial remedies. The conceptual seeds were present. Einstein’s 1905 paper, however, gave them a new status: he made the argument operational and precise, transforming what for Poincaré had been a philosophical or astronomical consideration into the foundation of a physical theory.
Miller’s account thus presents Poincaré as lingering in the background of Einstein’s reasoning, with his earlier reflections preparing the ground. Yet it was Einstein’s distinctive style of argument—his clarity, his insistence on operational definitions—that shaped the lasting framework of relativity. In Miller’s narrative, Poincaré is not erased but repositioned: less a background figure than a precursor whose presence accentuates, rather than diminishes, the originality of Einstein’s achievement.

Miller’s reading of Poincaré has the effect of retrofitting Einstein’s 1905 structure onto an earlier, and quite different, intellectual framework. He presents Poincaré as if he were already moving through the same three-step sequence that Einstein would later articulate: first, the empirical anchor of ether-drift experiments establishing the two-way constancy of light; second, a definitional stipulation of simultaneity in terms of synchronized clocks; and third, the synthesis whereby the velocity of light is elevated to the status of a universal postulate of kinematics. By casting Poincaré in this mold, Miller suggests that only the final “elevation” was lacking for Poincaré to achieve relativity.

Yet a closer examination of Poincaré’s writings complicates this neat alignment. His 1898 remarks on simultaneity were situated in the context of longitude determination in astronomy, not in the general kinematics of inertial frames. For Poincaré, this was not an operational definition in Einstein’s sense, but rather a fiction of ignorance — a methodological expedient justified by the smallness of possible discrepancies. Far from constituting a kinematical principle, simultaneity for Poincaré remained bound to astronomical practice and to the ether-based physics of his time.

Miller’s choice of notation further obscures the divergence. By importing Einstein’s later symbols and equations — $(t_A, t_B)$, and the synchronization formula \eqref{eq131} of the 1905 relativity paper — Miller narrates Poincaré’s reasoning in a language that Poincaré never used. This translation into Einstein’s idiom subtly erases the conceptual distance between the two thinkers. It creates the impression that the ether assumption in Poincaré’s framework was a minor leftover, rather than a structural commitment that fundamentally distinguished his approach from Einstein’s.

Seen in this light, the resemblance between Poincaré and Einstein is less a matter of parallel theoretical moves than of historiographical reconstruction. Poincaré’s conventions were pragmatic tools within an ether-based worldview, while Einstein’s definition of simultaneity was a constitutive stipulation that redefined the very architecture of kinematics. To project the latter back onto the former risks conflating two distinct intellectual enterprises, thereby narrowing the gap that was, in fact, decisive. 

In reality, the gulf is structural. Poincaré’s framework remains tied to the ether and treats simultaneity as conventional fiction, whereas Einstein fuses convention and empiricism into a universal principle of relativity.
Poincaré never crossed this threshold. In 1898, he recognized the conventionality of simultaneity and sketched the midpoint procedure, but only as a practical rule for determining longitude in astronomy. There is no invocation of $c$ as a universal constant, no fusion of convention with empirical invariance, and no kinematical framework built upon it. In 1900, he linked the midpoint procedure to Lorentz’s local time: observers in uniform motion, ignorant of their drift through the ether, assume equal one-way propagation times. But for Poincaré, this was a fiction: in truth, only the ether rest frame contained the “true time,” and only there was the one-way velocity isotropic. He accepted the two-way constancy of light speed as an empirical fact, yet he relegated one-way isotropy to an illusion produced by convention.

At no point did Poincaré promote the midpoint rule to a universal principle, nor did he fuse it with the empirical two-way invariance of light to declare $c$ a fundamental constant of kinematics. That final, audacious step—turning a convention into the bedrock of physical law—belongs uniquely to Einstein.

\subsection{Relativity of Simultaneity} \label{Sim}

Einstein demonstrated (using a thought experiment showing that simultaneity in one frame fails in the other) that the relativity of simultaneity follows qualitatively from the synchronization convention \eqref{eq114}.

Let us now consider two reference frames: System $K$ (Einstein calls it the “ruhendes System”). Light always propagates at speed $V$ relative to the system $K$.%
\footnote{At the time, most readers were still steeped in the ether framework of Lorentz and his contemporaries. To address them, Einstein retained the 19th-century jargon of a "stationary" system. Crucially, however, he stripped this term of its ontological weight: $K$ is not the ether, but simply the fiducial frame chosen for the derivation. 
Already in 1907, in his reply to Paul Ehrenfest, Einstein emphasized that relativity is not a "closed system" like Lorentz's ether theory but a heuristic principle \cite{Einstein07}. This distinction makes clear that in relativity any inertial system may be called "stationary," the designation being purely conventional.} We now consider a "moving" system $k$, in which a rod is at rest, while in the stationary system $K$ it moves uniformly with velocity $v$ along its axis.  
At the two ends of the rod are two clocks, $A$ and $B$. 
The question is whether these clocks remain synchronous according to the criterion \eqref{eq114}. 

Let a light signal be emitted from $A$ at time $t_A$, reflected at $B$ at time $t_B$, and received again at $A$ at time $t'_A$.  
We choose the $x$-axis of $K$ to be aligned with the direction of motion. 
Light propagates in $K$ with velocity $V$. 
The distance between the ends of the rod, measured at one instant of $K$, is denoted by $r_{AB}$. 
If the length of the rod at rest in $k$ (its proper length) is $L_0$, then in $K$ this length is contracted. However, in this stage, the contraction is implicit in the setup. The rod is moving in $K$, so its measured length is not $L_0$. But Einstein does not yet assert the exact formula. Only later, after introducing the Lorentz transformation, does Einstein make the length contraction explicit. 

At time $t_A$ in the "stationary" system $K$, a light signal leaves point $A$ (the left end of the rod). It travels towards point $B$, which is moving forward with velocity $v$. Because the rod is moving, the light has to “chase” the moving point $B$. So the relative speed of light with respect to the moving endpoint is $V-v$.

Let us first calculate the forward trip ($A \to B$). Suppose a light ray leaves point $A$ at time $t_A$ and reaches point $B$ at time $t_B$.
In the "stationary" system $K$, the light travels a distance:
\begin{equation} \label{A}
V(t_B - t_A).
\end{equation}
But endpoint $B$ is not standing still; during the same time interval, it has moved a distance:
\begin{equation} \label{B}
v(t_B - t_A).
\end{equation}
So the total distance the light must cover is:
\begin{equation} \label{C}
V(t_B - t_A) = r_{AB} + v(t_B - t_A).
\end{equation}
Rearranging this equation, we obtain:
\begin{equation} \label{D}
(V - v)(t_B - t_A) = r_{AB}, 
\end{equation}
Thus, the time it takes for the light to get from $A$ to $B$ is \cite{Einstein05}:%
\footnote{In many elementary derivations in relativity, one encounters algebraic expressions such as 
$\tfrac{1}{c-v}$ or $\tfrac{1}{c+v}$ when computing the times for light signals to catch up with, or return to, moving endpoints. 
At first glance, these denominators resemble the velocity--addition formulas of emission theories, where light would be assumed to propagate at $c \pm v$ relative to an observer. 
This superficial similarity has sometimes been confused. 
The crucial point, however, is that in special relativity such factors are not interpreted as the physical velocities of light. 
They arise only as algebraic consequences of enforcing Einstein’s postulate that light propagates at the invariant speed $c$ in the stationary system $K$ while the endpoints themselves move. 
Thus, the appearance of $c \pm v$ in the denominators should not be mistaken for a hidden appeal to emission theory.} 
\begin{equation} \label{eq115}
\boxed{t_B - t_A = \frac{r_{AB}}{V - v}.} 
\end{equation}

Next, Einstein lets the light be reflected at $B$ and travel back towards $A$. The endpoint $A$ is moving towards the light signal with velocity $v$. Thus, the relative speed of approach is $v+V$.

Let us calculate the return trip ($B \to A$). 
Now the light is reflected at $B$ and arrives again at $A$ at time $t'_A$.
In the system $K$, the light travels a distance:
\begin{equation} \label{A-1}
V(t'_A - t_B).
\end{equation}
But now point $A$ is moving \emph{towards} the light with velocity $v$, so the light has less distance to cover:
\begin{equation} \label{B-1}
r_{AB} - v(t'_A - t_B). \qquad \text{So:}
\end{equation}

\begin{equation} \label{C-1}
V(t'_A - t_B) = r_{AB} - v(t'_A - t_B).
\end{equation}
Rearranging this equation yields:
\begin{equation} \label{D-1}
(V + v)(t'_A - t_B) = r_{AB},
\end{equation}
The time for the return trip is therefore \cite{Einstein05}:

\begin{equation} \label{eq116}
\boxed{t'_A - t_B = \frac{r_{AB}}{V + v}.}    
\end{equation}

\vspace{2mm}
According to the synchronization definition \eqref{eq114}, one would require: $t_B - t_A = t'_A - t_B$. But as judged in the system $K$, equations \eqref{eq115}–\eqref{eq116} show that this condition is not satisfied for $v \neq 0$. We apply Einstein’s synchronization definition \eqref{eq114}, but we evaluate it in frame $K$ for the clocks riding on the rod. 

\noindent Substituting from equations \eqref{eq115} and \eqref{eq116}, we obtain:

\begin{equation} \label{E}
\frac{r_{AB}}{V - v} \;\neq\; \frac{r_{AB}}{V + v},    
\end{equation}
for $v \neq 0$. That is:
\begin{equation} \label{E-1}
t_B - t_A \;\neq\; t'_A - t_B.    
\end{equation}

\vspace{1mm}
Thus, equation \eqref{eq114} fails, and as judged in $K$, the moving clocks in $k$ do not satisfy the Einstein synchrony rule \eqref{eq114}. 
The mathematical derivation [\eqref{eq115}, \eqref{eq116}, \eqref{E}, and \eqref{E-1}] demonstrates that clocks synchronized in their own moving system appear out of synchronization when judged from the rest system $K$. 

Einstein generalizes this insight: the very notion of simultaneity is frame-dependent. He concludes: “We therefore see that we must not ascribe absolute meaning to the concept of simultaneity, but that two events which are simultaneous with respect to one coordinate system can no longer be regarded as simultaneous when viewed from a system that is moving relative to the first” \cite{Einstein05}. In other words, the system $K$ is moving relative to system $k$. Thus, Einstein’s conclusion amounts to saying that the rest system $K$ is moving with respect to the moving one $k$ just as much as the moving system $k$ is moving with respect to the rest one $K$. Einstein’s conclusion about simultaneity is the \textit{relativity of simultaneity}: simultaneity is not absolute, but depends on the state of motion of the observer. 

\subsection{Reconstruction and Presentation} \label{RP}

In his 1905 relativity paper, Einstein states that a light pulse leaves point $A$, travels to point $B$, and then returns to point $A$. He then immediately writes the travel times as equations \eqref{eq115} and \eqref{eq116}, but does not spell out the intermediate equations of motion or the coordinates of light and the endpoints. Instead, he appeals to his two principles:

\begin{enumerate}
    \item The principle of relativity (the moving system must be treated symmetrically).
    \item The constancy of the velocity of light $V$ in the system $K$.
\end{enumerate}

From those two alone, he argues that in the stationary system $K$, light moves at speed $V$, while the endpoints move at speed $v$. That is enough to justify the denominators in equations \eqref{eq115} and \eqref{eq116} $V \pm v$.

Thus, Einstein’s 1905 derivation of the relativity of simultaneity can be reconstructed with more detail than he chose to present in print (see the derivation in section \ref{Sim}). 
In the published paper, Einstein did not include the equations \eqref{A}, \eqref{B}, \eqref{C}, \eqref{D}, \eqref{A-1}, \eqref{B-1}, \eqref{C-1}, \eqref{D-1}, \eqref{E}, and \eqref{E-1}. Instead, in section \S 2, he reframed the issue operationally, using the measurement of a moving rod as his example. He distinguished two procedures: (a) measuring the rod with co-moving instruments, which by the principle of relativity must yield the same length $l$ as when the rod is stationary; and (b) measuring the rod in motion by means of stationary clocks in $K$, synchronized according to his earlier convention. The second procedure requires simultaneity in $K$, which is not the same as simultaneity in $k$. The result is that the length obtained by (b) is shorter than $l$, i.e.\ the length contraction effect.  

Why did Einstein adopt this presentation? Several reasons suggest themselves. First, rhetorical clarity and simplicity: explaining to readers the reasoning with rods and clocks rather than with equations, and Einstein’s operational framing gave the result vivid concreteness. Second, economy: he could compress the logical chain---synchronization, relativity of simultaneity, and contraction---into a single, tangible illustration. Third, didactic strategy: instead of leading the reader through a "light-chasing" calculation like the one reconstructed in section \ref{Sim}, he presented the relativity of simultaneity (and by extension also contraction) as following directly from the principle of relativity together with the light postulate.  

In Appendix \ref{AP}, I suggest an alternative algebraic derivation. This derivation is closely aligned with Einstein's operational definition of simultaneity \eqref{eq114}. 

\section{The Lorentz Transformation}

\subsection{Einstein's Path to Lorentz Transformation} \label{SL}

Einstein considered two Cartesian coordinate systems, each consisting of rigid rods and synchronized clocks \cite{Einstein05}.
System $K$ is "stationary" and $k$ moves with constant velocity $v$ along the common $X$-axis ($+x$ direction).
The $Y, Z$ axes are parallel in both systems. Clocks in each system are Einstein-synchronized [equation \eqref{eq114} or \eqref{eq1}]. $(x,y,z,t)$ are the $K$-coordinates of an event and $(\xi,\eta,\zeta,\tau)$ the $k$-coordinates of the same event. 

Einstein's operational definition of simultaneity, together with the requirement that light propagates with speed $c$, both lead to the Lorentz transformation \cite{Einstein05}.
We seek linear relations between the coordinates in $K$ $(x,y,z,t)$ and $k$ $(\xi,\eta,\zeta,\tau)$. We set \cite{Einstein05}:
\begin{equation}\label{eq27}
\boxed{x' \equiv x - v t.}
\end{equation}
At time $t=0$, the origins of $K$ and $k$ coincide. At later times, the origin of $k$ has a position in $K$: $x = vt$. Einstein temporarily introduces $x'$ as the Galilean distance between the moving origin of $k$ and the event in question, measured in $K$.%
\footnote{At this stage of the derivation, Einstein does not yet know the proper transformation law for $k$. So he introduces $x'$ as a temporary placeholder, a simple Galilean relative distance that is easy to compute in $K$. Later, after imposing the light postulate and linearity, he replaces these Galilean expressions with the full Lorentz–transformed coordinates $\xi, \eta, \zeta$.}

A point at rest in $k$ has constant $(x',y,z)$ when described from $K$.%
\footnote{\label{foot1} By \emph{homogeneity} and \emph{linearity}:
\begin{equation}
\xi = \alpha_{11} x + \alpha_{12} t, \qquad
\tau = \alpha_{21} x + \alpha_{22} t, \qquad
\eta = \alpha_{33} y, \qquad \zeta = \alpha_{44} z,
\end{equation}
with coefficients depending only on $v$. Since the $k$–origin has $x=vt$ in $K$ and $\xi=0$ in $k$, using equation \eqref{eq27} we may write: 
\begin{equation}\label{eq33}
\xi = A(v)\,(x - vt) = A(v)\,x', \qquad
\tau = B(v)\,x + C(v)\,t.
\end{equation}
In 1905, Einstein did not follow this route. However, the linearity in $t$ and $x$ is not an ansatz of convenience. It follows from the homogeneity of space and time, and the coefficient relations are pinned down by the light postulate and the reciprocity required by the relativity principle.}
Consider a light signal that leaves the $k$–origin at $\tau_0$, reflects at $x'$ at $\tau_1$,
and returns at $\tau_2$. Einstein synchronization in $k$ [equation \eqref{eq1}] requires \cite{Einstein05}:
\begin{equation}\label{eq30}
\boxed{\frac{\tau_0 + \tau_2}{2} = \tau_1.}
\end{equation}
Einstein expressed this in $(x',y,z,t)$ of $K$ using the light postulate:
\begin{equation} \label{eq71}
t_1 - t = \frac{x'}{c - v}, \qquad
t_2 - t_1 = \frac{x'}{c + v}.
\end{equation}
So the sequence of times in $K$ is:
\emph{Emission event in $K$}: The light ray is emitted from the origin of $k$. At the emission instant, the origins of $K$ and $k$ coincide.
So the emission event in $K$ is simply: $(0,0,0,t)$. 
\emph{Reflection event at $K$}: The ray travels with velocity $c$. It arrives at the point $x'$ of the $x$-axis of $k$ at time $\tau_1$. In $K$, since the point $x'$ is moving with velocity $v$, the ray meets it at [using equation \eqref{eq71}]: 
$(x',0,0,\,t + \frac{x'}{c-v})$. 
\emph{Return event at $K$}: The ray returns at time $\tau_2$ to the origin of $k$ with velocity $c$, while the origin itself moves with velocity $v$. In $K$, this is the event [using equation \eqref{eq71}]:
$(0,0,0,\,t + \frac{x'}{c-v} + \frac{x'}{c+v})$.
Einstein then replaces $\tau_0, \tau_1, \tau_2$ with the function $\tau(x', y, z, t)$ , i.e. the clock reading of the system $k$ at the event with coordinates $(x, y, z, t)$ at $K$, expressed in terms of the coordinate $x'$ [equation \eqref{eq27}]. 
Thus: 

\begin{equation} \label{eq135}
\tau_0=\tau(0,0,0,t), \\\   \tau_1=\tau(x',0,0,\,t + \frac{x'}{c-v}),  \\\ \tau_2=\tau(0,0,0,\,t + \frac{x'}{c-v} + \frac{x'}{c+v}).
\end{equation}

\noindent Einstein now substitutes equation \eqref{eq135} into the synchronization condition \eqref{eq30} written in terms of $k$'s time labels. Thus, he converts equation \eqref{eq30} into a functional equation for $\tau(x, y, z, t)$. So, plugging equation \eqref{eq135} into \eqref{eq30} gives \cite{Einstein05}: 
\begin{equation}\label{eq31}
\boxed{\frac{1}{2}\,\tau(0,0,0,t)
+ \frac{1}{2}\,\tau\!\left(0,0,0,\,t + \frac{x'}{c-v} + \frac{x'}{c+v}\right)
= \tau\!\left(x',0,0,\,t + \frac{x'}{c-v}\right).}
\end{equation}
Now he uses Taylor expansion to first order in $x'$. 

\noindent We treat $x'$ as infinitesimal and expand $\tau$ about $(0,0,0,t)$ to first order:
\begin{align}
\tau\!\left(0,0,0,\,t+\frac{x'}{c-v}+\frac{x'}{c+v}\right)
&\approx \tau(0,0,0,t)
+ \frac{\partial \tau}{\partial t}\!\left(\frac{x'}{c-v}+\frac{x'}{c+v}\right),\\
\tau\!\left(x',0,0,\,t+\frac{x'}{c-v}\right)
&\approx \tau(0,0,0,t)
+ \frac{\partial \tau}{\partial x'}\,x'
+ \frac{\partial \tau}{\partial t}\,\frac{x'}{c-v}.
\end{align}
We insert these into the synchronization equation \eqref{eq31}; the common $\tau(0,0,0,t)$ cancels, leaving \cite{Einstein05}:
\begin{equation}
\boxed{\frac{1}{2}\,\frac{\partial \tau}{\partial t}
  \left(\frac{x'}{c-v} + \frac{x'}{c+v}\right)
= \frac{\partial \tau}{\partial x'}\,x'
+ \frac{\partial \tau}{\partial t}\,\frac{x'}{c-v}.}
\end{equation}
Dividing by $x'$ and rearranging:
\begin{equation}
\frac{\partial \tau}{\partial x'}
= \frac{\partial \tau}{\partial t}
  \left[\tfrac{1}{2}\!\left(\frac{1}{c-v} + \frac{1}{c+v}\right)
   - \frac{1}{c-v}\right]. 
\end{equation}
Since:%
\footnote{The identity:
\begin{equation}
\frac{1}{2}\left(\frac{1}{c-v} + \frac{1}{c+v}\right) 
= \frac{1}{2} \cdot \frac{(c+v) + (c-v)}{(c-v)(c+v)} =
\frac{1}{2} \cdot \frac{2c}{c^2 - v^2} = \frac{c}{c^2 - v^2}.
\end{equation}}
\begin{equation} \label{eq74} 
\frac{1}{2}\!\left(\frac{1}{c-v} + \frac{1}{c+v}\right)
= \frac{c}{c^2 - v^2},
\qquad
\frac{c}{c^2-v^2} - \frac{1}{c-v}
= -\,\frac{v}{c^2-v^2},
\end{equation}
Einstein obtained the partial differential equation (PDE) \cite{Einstein05}:
\begin{equation} \label{eq65}
\boxed{\frac{\partial \tau}{\partial x'}
+ \frac{v}{\,c^2 - v^2\,}\,\frac{\partial \tau}{\partial t} = 0,
\qquad
\frac{\partial \tau}{\partial y}=0, \qquad \frac{\partial \tau}{\partial z}=0.}
\end{equation}
Now, we treat $t$ as fixed and integrate the PDE \eqref{eq65} in $x'$: 
We rearrange the equation as:
\begin{equation}
\frac{\partial \tau}{\partial x'} 
= -\frac{v}{c^2-v^2}\,\frac{\partial \tau}{\partial t}.
\end{equation}
That means $\tau$ is constant along straight lines with slope:
\begin{equation} \label{eq66}
\frac{\partial t}{\partial x'} = -\frac{v}{c^2-v^2}.
\end{equation}
Integrating equation \eqref{eq66}:
\begin{equation} \label{eq67}
\int \, \partial t = - \, \int \, \frac{v}{c^2-v^2}\, \partial x' = t- \, \frac{v}{c^2-v^2} \, x' = const. 
\end{equation}
So, $\tau$ can only depend on the combination:
\begin{equation} \label{eq68}
T \, = \, t - \frac{v}{c^2-v^2}\,x'.
\end{equation}
Thus, the general solution is:
\begin{equation} \label{eq75}
\tau(x',t) \, = F\!\left(t - \frac{v}{c^2-v^2}\,x'\right),
\end{equation}
where $F$ is an arbitrary differentiable function.

\noindent The PDE [\eqref{eq65}] tells us that $\tau$ must depend only on the single variable \eqref{eq68}. Now, if we further impose linearity in $(x',t)$, that means that $\tau$ must be a linear function of the single variable \eqref{eq68}. So we choose:

\begin{equation} \label{eq76}
F(T) = \, a(v) T, 
\end{equation}
giving \cite{Einstein05}:

\begin{equation}\label{eq32}
\boxed{\tau = a(v)\!\left(t - \frac{v}{\,c^2 - v^2\,}\,x'\right),}
\end{equation}
with $a(v)$ depending only on $v$.

\noindent We eliminate $x'$ in equation \eqref{eq32} via \eqref{eq27}:
\begin{equation}
\tau = a(v)\!\left(t - \frac{v}{c^2-v^2}(x - v t)\right),   
\end{equation}
and distribute the factor $\frac{v}{c^2-v^2}$ over $(x - v t)$:
\begin{equation}
\tau =a(v)\!\left(t - \frac{v}{c^2-v^2}x + \frac{v^2}{c^2-v^2}t\right).    
\end{equation}
Then we group the $t$ terms:
\begin{equation}
\tau = a(v)\!\left(\left(1+\frac{v^2}{c^2-v^2}\right)t - \frac{v}{c^2-v^2}x\right),    
\end{equation}
and combine the coefficients of $t$:
\begin{equation} \label{eq81}
1+\frac{v^2}{c^2-v^2}= \frac{c^2-v^2}{c^2-v^2} + \frac{v^2}{c^2-v^2}=\frac{c^2}{c^2-v^2}. \qquad \text{Hence:}   
\end{equation}

\begin{equation} \label{eq45}
\tau = a(v)\!\left(\frac{c^2}{c^2-v^2}\,t - \frac{v}{c^2-v^2}\,x\right).     
\end{equation}
Now we factor $\frac{c^2}{c^2-v^2}$ out of the bracket by writing the second term with a common factor:  
\begin{equation}
\tau = a(v)\,\frac{c^2}{c^2-v^2}\!\left(t - \frac{v}{c^2}x\right).  \quad \text{We use:}  
\end{equation}

\begin{equation} \label{eq53}
\gamma^2 \equiv \frac{1}{1-v^{2}/c^{2}}= \frac{c^2}{c^2-v^2},    \quad \text{to get:}    
\end{equation}

\begin{equation} \label{eq50}
\tau= a(v)\,\gamma^{2}\!\left(t - \frac{v}{c^{2}}x\right). 
\end{equation}
Thus, we derived that in $k$, the $\tau$ time coordinate is expressed linearly in $x, t$.

Einstein’s method for obtaining $\xi, \eta, \zeta$ is parallel. He uses the light postulate again, now requiring that light rays move with velocity $c$ in $k$ as well. He considers light emitted from the origin of $k$ at $\tau=0$ along the $\xi-$axis, $\eta-$axis and $\zeta-$axis. By enforcing that these light rays satisfy $\xi = \pm c\tau, \, \eta = \pm c \tau, \, \zeta = \pm c \tau$, in $k$, he extracts the transformation laws for $\xi, \eta, \zeta$.

\noindent \emph{For $\xi$}: Einstein considers a light ray moving in the positive $\xi$-direction in $k$. In the system $k$, such a ray must satisfy $\xi=c\tau$.  
Recall that $x'$ is a Galilean-relative displacement used as an intermediate variable: the distance in $K$ between the origin of $k$ at $x=vt$ and the event at $x$ [equation \eqref{eq27}].
Multiplying equation \eqref{eq32} by $c$ gives the form for $\xi$ \cite{Einstein05}:
\begin{equation} \label{eq47}
\boxed{\xi = a(v)c\left( t \;-\; \frac{v}{c^2 - v^2}\,x' \right).}
\end{equation}
This equation is consistent with the condition $\xi=c\tau$ for light moving along the $\xi-$axis.

\noindent In $K$, the ray along the $\xi$-axis satisfies \cite{Einstein05}:

\begin{equation} \label{eq80}
\boxed{t \;=\; \frac{x'}{\,c-v\,}.}
\end{equation}
We insert this equation into \eqref{eq47}:
\begin{equation} \label{eq48}
\xi \, =a(v)\,c\!\left(\frac{x'}{c-v}-\frac{v}{c^{2}-v^{2}}\,x'\right)
= a(v)\,c\,x'\!\left(\frac{1}{c-v}-\frac{v}{c^{2}-v^{2}}\right).    
\end{equation}
Now, we simplify the term:
\begin{equation} \label{eq46}
\frac{1}{c-v}-\frac{v}{c^{2}-v^{2}}.    
\end{equation}
in the equation \eqref{eq48}. First, we multiply the numerator and denominator by $c+v$, so the first term becomes:
\begin{equation} \label{eq55}
\frac{1}{c-v}=\frac{1}{c-v} \cdot \frac{c+v}{c+v} =\frac{c+v}{(c-v)(c+v)}= \frac{c+v}{c^2-v^2}.   
\end{equation}
We subtract [equation \eqref{eq46}]: 
\begin{equation}
\frac{c+v}{c^2-v^2}-\frac{v}{c^{2}-v^{2}}= \frac{c}{c^2-v^2},   
\end{equation}
and substitute this into equation \eqref{eq48} \cite{Einstein05}:
\begin{equation} \label{eq52}
\boxed{\xi=a(v)\,\frac{c^{2}}{c^{2}-v^{2}}\,x'} 
= a(v)\,\gamma^{2}\,x', \qquad \xi \;=\; a(v)\,\gamma^{2}\,(x - v t). 
\end{equation}

For a ray along the $y$-axis, $x' = 0$. So equation \eqref{eq47} reduces to: $\eta = a(v)\,\tau$. 
Einstein already multiplies $\tau$ by $c$ (because $\eta=c\tau$), So we get: 
\begin{equation} \label{eq49}
\eta = c\,\tau = a(v)\,c\,t.
\end{equation}
Thus, in the system $k$, a light ray that is purely transverse travels along the $\eta$-axis:
\begin{equation}
\xi = 0, \qquad \eta = c\tau, \qquad \zeta = 0.
\end{equation}
In the system $K$, every light signal belongs to the expanding sphere:
\begin{equation} \label{eq79}
x^2 + y^2 + z^2 = c^2 t^2.
\end{equation}
Since the origin of $k$ advances in the $x$-direction:
\begin{equation}
x = vt,
\end{equation}
the condition $\xi = 0$ translates, in $K$, into the requirement that the ray’s $x$-coordinate equals $vt$. Substituting this into the light-sphere equation \eqref{eq79} gives:
\begin{equation}
(vt)^2 + y^2 = c^2 t^2.
\end{equation}
Solving for $t$ in terms of $y$ yields in $K$ \cite{Einstein05}:

\begin{equation} \label{eq70}
\boxed{t = \frac{y}{\sqrt{c^{2}-v^{2}}},}
\end{equation}
which when inserted into equation \eqref{eq49} gives \cite{Einstein05}:
\begin{equation}
\boxed{\eta = a(v)\,\frac{c}{\sqrt{c^{2}-v^{2}}}\,y.} 
\end{equation}
Analogously for the $\zeta$-axis \cite{Einstein05}:
\begin{equation}
\boxed{\zeta = a(v)\,\frac{c}{\sqrt{c^{2}-v^{2}}}\,z.} 
\end{equation}
\noindent  These become: 
\begin{equation} \label{eq51}
\eta=a(v)\,\gamma\,y \qquad \zeta=a(v)\,\gamma\,z.    
\end{equation}

Finally, once Einstein had worked out each coordinate separately --- equations \eqref{eq50}, \eqref{eq52}, and \eqref{eq51} --- under the constraints of the light postulate and linearity, he gathered them into the full set of transformations. 
Since $a(v)$ is still arbitrary, he absorbed one factor of $\gamma$ into it, and defined a new function:

\begin{equation} \label{eq77}
\phi(v)=a(v)\gamma,    
\end{equation}
yielding the final form of the space and time transformation \cite{Einstein05}:\footnote{\label{Foot2} Imposing the light condition along $+\xi$ ($\xi=c\tau$ for a light ray in $k$)
fixes the linear coefficients in \eqref{eq33} to (Einstein did not follow this path):

\begin{equation}
A(v)=\phi(v)\gamma,\qquad B(v)=-\,\phi(v)\gamma\,\frac{v}{c^{2}},\qquad C(v)=\phi(v)\gamma.    
\end{equation}
These coefficients are not merely algebraic conveniences; they represent the physical postulates that Einstein spelled out — homogeneity, isotropy, and relativity.} 

\begin{equation} \label{eq9}
\tau = \phi(v)\,\gamma\!\left(t - \frac{v}{c^2}\,x \right), \quad \xi = \phi(v)\,\gamma\,(x - v t), \quad \eta = \phi(v)\,y, \quad \zeta = \phi(v)\,z,   
\end{equation} 
\begin{equation} \label{eq34} 
\text{with the square root of equation \eqref{eq53}:} \qquad \gamma \equiv \frac{1}{\sqrt{1-v^{2}/c^{2}}}
= \frac{c}{\sqrt{c^{2}-v^{2}}}.     
\end{equation}

\subsection{Compatibility Check of Relativity and Light Postulate} \label{SW}

To conclude that the same numerical constant $c$ holds in \emph{every} inertial frame, Einstein added the relativity principle. 
A spherical wave emitted at $t=\tau=0$ satisfies in $K$ \cite{Einstein05}:
\begin{equation} \label{eq10}
\boxed{x^2 + y^2 + z^2 = c^2 t^2}.
\end{equation}
Transforming with the Lorentz transformation gives \cite{Einstein05}:
\begin{equation} \label{eq10-1}
\boxed{\xi^2 + \eta^2 + \zeta^2 = c^2 \tau^2},
\end{equation}
showing that in $k$ the wave is also spherical with speed $c$. 

The spherical-wave transformation is a consistency check.
From the provisional transformation \eqref{eq9}, one finds:
\begin{equation} \label{eq11}
\xi^2+\eta^2+\zeta^2-c^2\tau^2
=\phi(v)^2\Big[x^2+y^2+z^2-c^2 t^2\Big].
\end{equation}
Hence, a spherical wave in $K$ [equation \eqref{eq10}] maps to a spherical wave in $k$ [equation \eqref{eq11}]
for any $\phi(v)$. Thus, if the wavefront in $K$ satisfies \eqref{eq10}, then in $k$ we also get \eqref{eq10-1}, regardless of $\phi(v)$.

This shows that the two postulates are not contradictory. Light remains spherical in every frame, \emph{even before $\phi(v)$ is fixed}. That is why Einstein calls it a compatibility or consistency check. So this step checks the compatibility of his two postulates, but it does not fix $\phi(v)$. 
At this stage, Einstein is still deriving the transformation, not assuming it.
So he allows the most general linear form consistent with homogeneity and isotropy, introducing an undetermined factor $\phi(v)$. He cannot yet assume the final Lorentz form (with $\phi(v)=1$) \eqref{eq12} because that would beg the question. Hence, he checks the spherical wave law with the ansatz \eqref{eq9}, to demonstrate that the two postulates are compatible before fixing $\phi(v)$.

\subsection{\texorpdfstring{Fixing $\phi(v)=1$}{Fixing phi(v)=1}}

Subsequently, Einstein fixed $\phi(v)$. Einstein’s 1905 argument for fixing $\phi(v)$ rests on physical reciprocity (inverse transformation has the same form) and transverse symmetry (no contraction perpendicular to the motion):

\emph{Step 1. Reciprocity condition:} Einstein introduced a third system $K'$ moving with velocity $-v$ relative to $k$ (parallel to $X$).
A double application of the transformation (from $K$ to $k$ and then $k$ to $K'$) gives \cite{Einstein05}:
\begin{equation} \label{eq36}
\boxed{\phi(v)\,\phi(-v) = 1.}
\end{equation}
Thus, applying forward and backward transformations brings us back to the identity.

\emph{Step 2. Symmetry condition}: Now consider a rod of length $l$ at rest along the $\eta$-axis of $k$ (so it moves perpendicular to its axis in $K$).
Its endpoints have $K$-coordinates $(x_1,y_1,z_1)=(vt,\,l/\phi(v),\,0)$ and $(x_2,y_2,z_2)=(vt,\,0,\,0)$, hence its length measured in $K$ is $l/\phi(v)$.
By symmetry this must be unchanged under $v\!\to\!-v$, so \cite{Einstein05}:
\begin{equation} \label{eq37}
\boxed{\phi(v)=\phi(-v).}
\end{equation}

\emph{Step 3. Combining the two:} Einstein substituted the symmetry condition \eqref{eq37} into the reciprocity condition \eqref{eq36} \cite{Einstein05}:
\begin{equation}
\phi(v)\,\phi(-v) = \phi(v)\,\phi(v) = \big(\phi(v)\big)^2 = 1. \qquad \text{So:}
\end{equation}

\begin{equation} \label{eq39}
\phi(v)^2 = 1 \;\;\;\Rightarrow\;\;\; \phi(v) = \pm 1.
\end{equation}

\noindent At this stage, Einstein rules out the negative sign $\phi(v)=-1$ because it would flip time order and contradict the physical requirement that $\tau$ increase with $t$. In other words, the transformation would flip the signs of time/space in an unphysical way, and it would reverse orientation and simultaneity conventions. 
Thus, Einstein finally chose: 

\begin{equation} \label{eq40}
\phi(v)=1.
\end{equation}

\noindent The final form of the Lorentz transformation  is therefore equation \eqref{eq2} \cite{Einstein05}: 
\begin{equation} \label{eq12}
\boxed{\tau = \gamma\!\left( t - \frac{v}{c^2}\,x \right), \qquad
\xi = \gamma\,(x - v t), \qquad
\eta = y, \qquad
\zeta = z,}
\end{equation}

\noindent The Lorentz transformation in standard form:
\begin{equation} \label{eq2}
\,x'=\gamma(x-vt),\quad t'=\gamma\!\left(t-\frac{v}{c^2}x\right),\quad y'=y,\ z'=z.\, \quad
\gamma=\frac{1}{\sqrt{1-\frac{v^2}{c^2}}}.
\end{equation}

\subsection{Einstein's Hidden Algebraic Eliminations}

In 1905, Einstein did not adopt the straightforward route that now appears so natural (see footnotes \ref{foot1} and \ref{Foot2}). Instead, he grounded his derivation of the Lorentz transformation in concrete operational definitions: (1) synchronization by light signals, and (2) the measurement of simultaneity and length with physical clocks and rods. 

Einstein's derivation is minimalist, operational, and physically motivated. His mathematical tools are mostly algebra and simple functional arguments. He employs linearity, symmetry, and invariance reasoning rather than relying on explicit, computationally intensive methods. The algebraic scaffolding behind the printed derivation in the \emph{Annalen der Physik} paper (the boxed equations) is hidden; the published text is stripped down to the essentials.

In the relativity paper, Einstein keeps the algebra as light as possible. His goal is to convince physicists that relativity can be derived from two simple principles (the principle of relativity and the postulate of the velocity of light) and basic operational definitions, rather than from complex electrodynamics. He bypasses the machinery of Lorentz's electron theory, and his mathematics is intentionally light to serve that rhetorical aim.
His “lack” of mathematical sophistication is a deliberate rhetorical and philosophical move: the mathematics serves the principles, not the other way around. 

Yet behind the published derivation of the Lorentz transformation lies a sequence of algebraic eliminations and constraints that Einstein compressed and omitted. Below, I reconstruct these steps explicitly, with the numbering of the equations matching the equations in my detailed derivation in section \ref{SL}:

1) \emph{Synchronization and the PDE for \texorpdfstring{$\tau$}{tau}}: Synchronization in $k$ requires equation \eqref{eq30}. Using the light postulate in $K$, the corresponding time intervals are given by equation \eqref{eq71}, leading to equation \eqref{eq31}. Expanding to first order in $x'$ and canceling the common term $\tau(0,0,0,t)$ yields the PDE \eqref{eq65}. This step performs two eliminations at once: discarding higher-order terms and removing the redundant $\tau$ value. At the same time, the forward and backward light-travel times, $\tfrac{1}{c-v}$ and $\tfrac{1}{c+v}$, combine into their symmetric average [equation \eqref{eq74}]. Algebraically, this is a neat simplification, but its real significance is physical: the averaging embodies Einstein’s synchronization rule, namely that light takes the same time to go out and back. What appears to be an algebraic trick is, in fact, the mathematical expression of the round-trip constancy of $c$, the principle that ensures consistency of synchronization in both frames.

2) \emph{Reducing a $2$-variable dependence to $1$ invariant combination}: 
At the outset, Einstein defines $k$’s time coordinate as a general function $\tau(x',t)$, depending on the Galilean-relative displacement $x'$ \eqref{eq27} and $K$’s time $t$. Imposing the synchronization requirement \eqref{eq30} [\eqref{eq1}] and the constancy of $c$ yields the differential relation \eqref{eq65}, a first-order linear PDE that ties the two variables together. It states that $\tau$ cannot vary independently in $x'$ and $t$: any change in one must be accompanied by a corresponding change in the other. The characteristic curves of this PDE are straight lines in the $(x',t)$-plane \eqref{eq66}, along which $\tau$ remains constant.

The general solution \eqref{eq75} shows that $\tau$ depends only on a single invariant combination of $t$ and $x'$, equation \eqref{eq68}. 
Thus, a two-variable dependence collapses into one. All solutions of the PDE reduce to functions of $T$. 

Einstein appeals to the homogeneity of space and time and the linearity of inertial transformations. If the laws of physics are the same everywhere and at every time, then the transformation between coordinates cannot involve nonlinear distortions of $t$ and $x$. That restriction forces $F$ to be a linear function of its argument. So Einstein reduces $F$ to equation \eqref{eq76}, yielding Einstein’s expression \eqref{eq32}.

All solutions of the PDE collapse onto a single-variable dependence on $T$. Physically, this means that only the specific combination $T$ is relevant for defining time in $k$. The PDE reduces the degrees of freedom: $\tau$ is no longer an arbitrary surface over $(x',t)$ but a single-variable dependence lifted into two dimensions. Since $T$ mixes spatial and temporal coordinates, simultaneity in $k$ depends on both $t$ and $x'$. Events simultaneous in $K$ need not be simultaneous in $k$, because time in $k$ is inseparably tied to spatial position in $K$. This interweaving of space and time abolishes absolute simultaneity, preparing the ground for the Lorentz transformation.

3) \emph{The hidden algebra behind the $\xi$-transformation}: At this stage Einstein already had a candidate form for $\tau$, equation \eqref{eq32}. Imposing the light condition in $k$ that a ray along the $\xi$-axis must satisfy $\xi = c \tau$, and substituting the form of $\tau$ \eqref{eq32}, gives equation \eqref{eq47}. In $K$, a light ray along the $\xi$-axis obeys equation \eqref{eq80}. Substituting this into equation \eqref{eq47} and simplifying yields equation \eqref{eq48}. Here, Einstein employs a hidden algebraic maneuver. The parenthetical expression in equation \eqref{eq48} is not obvious at first glance. To reduce it, he rewrites $\tfrac{1}{c-v}$ with a common denominator \eqref{eq55}, so that subtracting $\tfrac{v}{c^2 - v^2}$ directly gives $\tfrac{c}{c^2-v^2}$. This partial-fraction trick converts a messy, asymmetric term into a clean, symmetric factor, reflecting the two-way light condition once again. Substituting back, we arrive at equation \eqref{eq52} [using equation \eqref{eq27}]. Einstein introduces this neat algebraic simplification that appears almost hidden if one reads quickly. 

\subsection{Did Einstein work backwards?} 

Miller has argued that Einstein worked backwards, already knowing the Lorentz transformation and then arranging his derivation in 1905 to reproduce it (he "knew beforehand the special portion of the relativistic transformations and an approximate version of the correct time coordinate"). Miller bases this claim on the apparently sudden introduction of the factor $a(v)\sqrt{1-v^{2}/c^{2}}$ [equation \eqref{eq50}] and on the compressed algebra of the 1905 paper, which to him suggests retrofitting rather than genuine derivation \cite{Miller}.

This interpretation, however, is misleading. The 1905 paper is famously austere, presenting only the barest outline of the argument. When reconstructed in full detail, as in equations \eqref{eq32}--\eqref{eq50}, the derivation requires substantial algebraic manipulations such as forming common denominators, regrouping terms, and partial-fraction expansions. Einstein’s suppression of these steps is entirely consistent with his later style in general relativity and gravitational waves, where the scaffolding is likewise invisible \cite{Weinstein}. The appearance of elegance should not be mistaken for foreknowledge.

The oft-criticized step of replacing $a(v)$ with $\phi(v)\sqrt{1-v^{2}/c^{2}}$ [equation \eqref{eq77}] is not evidence of reverse engineering Lorentz but a careful operational progression. It is simply an algebraic reparametrization: one absorbs a factor of $\gamma$ to simplify notation. The decisive constraints that fix $\phi(v)$ to unity come later. 
Miller (and others) argue that since Einstein had studied Lorentz’s 1895 \emph{Versuch} \cite{Lorentz1895}, he must have already known that Lorentz’s transformations worked only if the scale factor $\phi(v)=1$. Thus, in 1905, when Einstein introduced $\phi(v)$ as an arbitrary function and then “discovered” it equals unity, Miller claims he was reenacting what Lorentz had already done, not deriving it independently \cite{Miller}. 

Crucially, Einstein did not initially fix the transformation coefficients. Instead, he introduced an undetermined scale factor $a(v)$, carried through several stages of the calculation. The synchronization condition \eqref{eq30} and the light postulate yield a partial differential equation \eqref{eq65}, whose general solution is equation \eqref{eq75},
with $F$ arbitrary. Imposing linearity then restricts this to equation \eqref{eq76} but still leaves $a(v)$ undetermined. Only by enforcing the light postulate in multiple directions, and subsequently applying reciprocity and continuity at $v=0$, does Einstein arrive at the final Lorentz transformation with equation \eqref{eq40} $\phi(v)=1$. Thus, if Einstein had imported Lorentz’s result, he would never have left $\phi(v)$ undetermined in the equations \eqref{eq9}. Carrying $\phi(v)$ through half the derivation until reciprocity and continuity are applied to fix it to unity shows that Einstein was not borrowing, but rather re-deriving the transformation within a new conceptual framework. This makes his derivation robust; the principle of relativity itself, not Lorentz’s theory of the electron, sets $\phi(v)=1$.

Thus, when reconstructed carefully, Einstein’s method reveals a stepwise progression from synchronization to the Lorentz transformation, rather than the retrofitted argument Miller supposes. The presence of undetermined functions throughout demonstrates that Einstein did not begin with the Lorentz transformation, but arrived at it systematically by applying his postulates. 

\subsection{Length Contraction and Time Dilation} \label{LC}

Einstein in 1905 derived length contraction and time dilation explicitly from the Lorentz transformations \eqref{eq12}, but \emph{he did not derive the relativity of simultaneity in the same way} (see the explanation in section \ref{RP}). 

Einstein first takes a rigid sphere at rest in $k$, transforms it to $K$, and shows that it becomes an ellipsoid, i.e., the $x$-axis is contracted by $\sqrt{1-v^2/c^2}$ (length contraction).
He then places a clock at the origin of $k$ and transforms its readings into $K$, showing that the moving clock runs slow by the same factor (time dilation).
These are both direct kinematic consequences of the transformation equations. 

Subsequently, Einstein derived the clock paradox from the time dilation relation \cite{Einstein05}:

\begin{equation} \label{eq112}
\tau = t \sqrt{1 - \frac{v^2}{c^2}},  \qquad \text{exact and valid for all} \quad v<c.
\end{equation}
where $t$ is the coordinate time measured in system $K$, $\tau$ is the time interval measured by the moving clock, $v$ is the velocity of the moving clock relative to $K$, and $c$ is the speed of light.

Einstein then considers two clocks: Clock $A$ at point $A$, moving with velocity $v$ from $A$ to $B$. Clock $B$ is stationary at point $B$, synchronized with $A$ at the start (in the system $K$). From the time dilation relation \eqref{eq112}, When the moving clock $A$ reaches point $B$, less time has elapsed on it compared to the stationary clock $B$ \cite{Einstein05}. 

Mathematically, Einstein expands the time dilation formula to second order. He wanted to quantify the clock lag up to order $\frac{v^2}{c^2}$. So, he expands the square root $\sqrt{1 - \frac{v^2}{c^2}}$ using the binomial (Taylor) series approximation in $\frac{v}{c}$: 
\begin{equation}
\sqrt{1 - x} = 1 - \tfrac{1}{2}x - \tfrac{1}{8}x^2 - \tfrac{1}{16}x^3 - \cdots,     \quad \text{valid for } |x| < 1. \quad \text{Here:} \quad x = \frac{v^2}{c^2}. 
\end{equation}

\begin{equation}
\text{Thus:} \qquad \sqrt{1 - \frac{v^2}{c^2}}
\approx 1 - \tfrac{1}{2}\frac{v^2}{c^2} - \tfrac{1}{8}\frac{v^4}{c^4} - \cdots.
\end{equation}
Keeping only terms up to order $\mathcal{O}\!\left(\frac{v^2}{c^2}\right)$ (i.e., truncating after the quadratic term), Einstein obtained:
\begin{equation}
\sqrt{1 - \frac{v^2}{c^2}} \approx 1 - \tfrac{1}{2}\frac{v^2}{c^2}. 
\end{equation}
Therefore, the second–order Taylor (binomial) approximation of the formula \eqref{eq112} is:
\begin{equation} \label{eq113}
\boxed{\tau = t \sqrt{1 - \frac{v^2}{c^2}} \quad \rightarrow \quad t- \tau =\!\left[1-\sqrt{1-\frac{v^2}{c^2}}\right]t} 
\approx t \left( 1 - \tfrac{1}{2}\frac{v^2}{c^2} \right),
\end{equation}
which is valid only for small $\frac{v}{c}$ (non-relativistic regime). 

\noindent This shows that after travel time $t$, the moving clock $A$ lags behind the stationary clock by \cite{Einstein05}:

\begin{equation}
\Delta t \equiv t-\tau
\approx \boxed{\tfrac{1}{2}t\,\frac{v^2}{c^2}.}    
\end{equation}

According to Jean-Marc Ginoux's recent book, \emph{Poincaré, Einstein and the Discovery of Special Relativity: An End to the Controversy}, "... Einstein did indeed use Taylor’s series expansions in his demonstrations. However, in the case of 'Langevin’s twins paradox,' they can no longer be used, since Langevin considers a traveller moving at a speed very close to that of light" \cite{Gin}.

First, Einstein neither invoked ``twins'' nor described a paradox. Instead, he drew a direct corollary from the time--dilation law, applying it to closed trajectories and showing: (1) a moving clock traversing a polygonal path from $A$ to $B$ and back accumulates less time than a stationary clock; (2) the same holds along continuously curved paths; and (3) most generally, any closed journey results in the moving clock lagging behind the one that remained at rest \cite{Einstein05}.  

Second, the expression for the lag that appears in the 1905 paper is a low-velocity approximation of Einstein’s exact result. But Einstein had already derived the exact relation \eqref{eq112}; the Taylor expansion was introduced only to indicate the order of magnitude of the effect in practically relevant, slow-motion contexts (e.g., terrestrial clocks, astronomical motions). Its role was pedagogical, not foundational. The general conclusion---that a moving clock always accumulates less proper time upon reunion---follows directly from the exact formula \eqref{eq112}, valid at all velocities $v<c$. While the truncated series cannot be extrapolated to near-light speeds, Einstein’s reasoning never relied on it: the inequality $\tau < t$ for $v \neq 0$ holds without approximation.

\section{Relativity of Velocity Formulas}

\subsection{Einstein and Poincaré Side by Side}

This section places side by side the actual derivations in Einstein’s 1905 paper and in Poincaré’s writings—his private letter to Lorentz of May 1905 and his \emph{Rendiconti di Palermo} memoir of 1906 \cite{Poi05-2}. By contrast, Poincaré’s brief June 1905 note in \emph{Comptes Rendus} \cite{Poi05-1} contains no such derivations. The chronology speaks for itself: Einstein could not have had access to Poincaré’s private correspondence with Lorentz. As a patent clerk in Bern, he had no pipeline to Lorentz’s mail; those connections to Leiden came only later. And Poincaré’s detailed memoir of Palermo was published after Einstein’s June 1905 article. In short, the documents themselves close the door on any notion of borrowing.

On the Einstein side, I start from section \S 5 of the kinematical part \cite{Einstein05}, where the Lorentz transformations are used operationally to obtain the relativistic addition law of velocities. On the Poincaré side, I reconstruct the route from the correspondence and from “Sur la dynamique de l’électron” \cite{Poi05-2} where the Lorentz transformations are written with a dimensionless parameter and composed to exhibit group structure, including the condition that fixes the scale factor. In both reconstructions, the computations are kept at the level of what the authors print or clearly imply, without importing later formalisms.

I reconstruct a terse remark of Einstein’s—his note that successive parallel transformations “form a group”—into an explicit composition calculation. This closes the loop between the section \S 5 addition law and the transformation-composition viewpoint without changing the content of Einstein’s paper: identity and inverses are implicit, associativity is not discussed, and rotations are outside the scope. The point is limited and precise: even read as a kinematical check, the collinear case reproduces the same parameter \eqref{eq41} and the corresponding Lorentz factor, showing closure exactly where Einstein says it should appear.

The upshot is that there is less mystery here than recent polemics suggest. When one follows the printed steps, both authors arrive at the same formulas; what distinguishes them is where those formulas reside. In Poincaré’s hands, they are instruments inside an ether-anchored invariance program; in Einstein’s hands, they become the laws of a new kinematics. The present paper keeps to that concrete ground: it documents the derivations as given, makes the group-composition linkage explicit in the collinear case, and clarifies why “the same equation” can be, historically and conceptually, a different claim.

\subsection{The Relativistic Addition Law of Velocities} \label{RV}

In section §5 of his 1905 paper, Einstein used the Lorentz transformations to derive the relativistic velocity addition law. Consider two systems. System $K$ is "stationary" and $k$ moves with constant velocity $v$ along the common $X$-axis ($+x$ direction).%
\footnote{The designation of “stationary” or “moving” in Einstein’s 1905 paper is purely conventional. No inertial frame is privileged (so “stationary” does not mean “ether-rest”). If, instead, we declare $k$ to be “stationary” and $K$ to be “moving” with velocity $-v$ along the $x$-axis, we invert the transformation. The inverse Lorentz transformation is \eqref{eq12I}.}
The $Y$ and $Z$ axes are parallel in both systems. $(x,y,z,t)$ are the $K$-coordinates of an event and $(\xi,\eta,\zeta,\tau)$ the $k$-coordinates of the same event. His derivation involves transforming a straight-line motion in the system $k$ into the system $K$ and then reading off the velocity components. This yields the relativistic addition law of velocities.

Specifically, Einstein considered two inertial frames, system $K$ and another system $k$ moving with velocity $v$ along the $X$-axis of $K$.  
In $k$, a particle moves as \cite{Einstein05}:
\begin{equation} \label{eqM}
\boxed{\xi = w_\xi \tau, \qquad \eta = w_\eta \tau, \qquad \zeta = 0.}
\end{equation}
Using the Lorentz transformations \eqref{eq12}, Einstein wanted the inverse transformations, i.e., to express ($x, t$) in terms of ($\xi, \tau$).%
\footnote{First, we take the two coupled equations: 

\begin{equation} \label{X}
\xi = \gamma\,x - \gamma \, v t,    
\end{equation}
\begin{equation} \label{T}
\tau = \gamma \, t - \gamma \frac{v}{c^2}\,x.    
\end{equation}
Then we solve the equation \eqref{X} for $x$:

\begin{equation} \label{X-1}
\frac{\xi}{\gamma} = x - v t 
\quad \Rightarrow \quad 
x = \frac{\xi}{\gamma} + v t.
\end{equation}
We substitute equation \eqref{X-1} into equation \eqref{T}:

\begin{equation}
\tau= \gamma t - \gamma \frac{v}{c^2}\!\left(\frac{\xi}{\gamma} + v t\right) =  \gamma t - \frac{v}{c^2}\xi - \gamma \frac{v^2}{c^2} t =  \gamma t \!\left(1 - \frac{v^2}{c^2}\right) - \frac{v}{c^2}\xi. 
\end{equation}
But recall
\begin{equation}
1 - \frac{v^2}{c^2} = \frac{1}{\gamma^2}. \qquad \text{So:} \qquad  \tau = \frac{\gamma}{\gamma^2} t - \frac{v}{c^2}\xi 
= \frac{t}{\gamma} - \frac{v}{c^2}\xi.
\end{equation}
We solve for $t$: 

\begin{equation} \label{t}
\frac{t}{\gamma} = \tau + \frac{v}{c^2}\xi 
\quad \Rightarrow \quad
\boxed{t = \gamma \!\left(\tau + \frac{v}{c^2}\xi\right).}
\end{equation}
Now, from equation \eqref{X} we find $x$:

\begin{equation} \label{gX}
\gamma x =  \xi + \gamma v t \quad \Rightarrow \quad x = \frac{\xi}{\gamma} + v t.
\end{equation}
Now we substitute the $t$ \eqref{t} we just found into equation \eqref{gX}:

\begin{equation}
x= \frac{\xi}{\gamma} + v \gamma\!\left(\tau + \frac{v}{c^2}\xi\right) = \frac{\xi}{\gamma} + \gamma v \tau + \gamma \frac{v^2}{c^2}\xi.   
\end{equation}
We combine the $\xi$-terms:

\begin{equation}
\frac{\xi}{\gamma} + \gamma \frac{v^2}{c^2}\xi = 
\xi\!\left(\frac{1}{\gamma} + \gamma \frac{v^2}{c^2}\right).
\end{equation}
Since $\frac{1}{\gamma} + \gamma \frac{v^2}{c^2} = \gamma$, we finally find: 

\begin{equation}
x = \gamma \xi + \gamma v \tau \qquad \Rightarrow \qquad \boxed{x=\gamma(\xi + v \tau).}
\end{equation}}
The result is:

\begin{equation} \label{eq12I}
x = \gamma(\xi + v \tau), \qquad y = \eta, \qquad 
t = \gamma\!\left(\tau + \tfrac{v}{c^2}\xi\right),
\end{equation}
He substituted the particle’s motion \eqref{eqM} into equation \eqref{eq12I}:
\begin{equation}
x = \gamma(w_\xi + v)\tau, \qquad 
y = w_\eta \tau, \qquad 
t = \gamma\!\left(1 + \tfrac{v w_\xi}{c^2}\right)\tau.
\end{equation}

\noindent He divided $x$ by $t$ and $y$ by $t$ to obtain the components ($x,y,z$) in $K$ \cite{Einstein05}:
\begin{equation} \label{eq87}
\boxed{x = \frac{w_\xi + v}{1 + \tfrac{v w_\xi}{c^2}} \, t, \qquad
y = \frac{w_\eta}{\gamma\!\left(1 + \tfrac{v w_\xi} {c^2}\right)} \, t, \qquad 
z = 0.}
\end{equation}
Einstein notes that to first approximation, the old parallelogram law still works, but in the general case we need the corrected relativistic formula \eqref{eq87} \cite{Einstein05}.

\noindent Einstein then wants the magnitude of the resultant velocity.
The resultant velocity satisfies \cite{Einstein05}:
\begin{equation} \label{eq104}
\boxed{U^2= \left(\frac{dx}{dt}\right)^2  + \, \left(\frac{dy}{dt}\right)^2.}   
\end{equation} 
Now, he defines $w^2$, and introduces polar coordinates for $w$ \cite{Einstein05}:
\begin{equation}
\boxed{w^2= w_\xi^2 \,+ w_\eta^2 \qquad  \alpha = \tan^{-1} \, \left(\frac{w_\eta}{w_\xi}\right),} 
\end{equation}
where $\alpha$ is the angle between $v$ and $w$, and:
\begin{equation}
w_\xi = w \cos\alpha, \qquad w_\eta = w \sin\alpha.    
\end{equation}

\noindent Plugging these into equation \eqref{eq87}, we obtain:
\begin{equation}
\frac{x}{t} = \frac{v + w \cos\alpha}{1 + \tfrac{v w}{c^2}\cos\alpha}, 
\qquad
\frac{y}{t} = \frac{w \sin\alpha}{\gamma\left(1 + \tfrac{v w}{c^2}\cos\alpha\right)}=\frac{w \sin\alpha \sqrt{1 - v^2/c^2}}{1 + \tfrac{v w}{c^2}\cos\alpha}. 
\end{equation}
Now we compute equation \eqref{eq104}:

\begin{equation}
U^2 = \frac{(v + w \cos\alpha)^2}{\left(1 + \tfrac{v w}{c^2}\cos\alpha\right)^2}
   + \frac{(1 - v^2/c^2)(w \sin\alpha)^2}{\left(1 + \tfrac{v w}{c^2}\cos\alpha\right)^2}.   
\end{equation}
We combine terms in the numerator:
\begin{equation}
(v + w \cos\alpha)^2 + (1 - v^2/c^2)(w \sin\alpha)^2 
= v^2 + w^2 + 2 v w \cos\alpha - \frac{(v w \sin\alpha)^2}{c^2}. 
\end{equation}
Hence the resultant speed is \cite{Einstein05}:
\begin{equation} \label{eq42}
\boxed{U = \frac{\sqrt{v^2 + w^2 + 2 v w \cos\alpha - \left(\tfrac{v w \sin\alpha}{c}\right)^2}}{1 + \tfrac{v w}{c^2}\cos\alpha}.}
\end{equation}
For $\alpha = 0$, we have $\cos\alpha = 1$ and $\sin\alpha = 0$;  equation \eqref{eq42} reduced to:

\begin{equation}
U = \frac{\sqrt{v^2 + w^2 + 2 v w - 0}}{1 + \tfrac{v w}{c^2}\cdot 1}   = \frac{\sqrt{(v+w)^2}}{1 + \tfrac{v w}{c^2}} 
  = \frac{|v+w|}{1 + \tfrac{v w}{c^2}}.    
\end{equation}
If $w$ is along the $X$-axis (collinear, the same direction), $v,w \geq 0$, we take the positive root, so \cite{Einstein05}:

\begin{equation} \label{eq26}
\boxed{U = \frac{v+w}{1 + \tfrac{vw}{c^2}},}
\end{equation}
which reduces to the Galilean addition law of velocities when $v<<c$.%
\footnote{In Newtonian mechanics, if we have two velocities $\vec{v}$ and $\vec{w}$, the resultant is given by the parallelogram rule (vector addition): $\vec{U}=\vec{v}+\vec{w}$. But in equation \eqref{eq26}, because of the denominator $1 + \tfrac{v w}{c^2}$, velocities do not add like Newtonian vectors. Only in the limit $v << c$ (for small speeds compared to $c$, i.e., to first-order approximation) does the denominator become $\sim 1$, and then we recover the parallelogram rule.}
Equation \eqref{eq26} guarantees that the result of adding two subluminal velocities is still subluminal, and preserves the invariance of the speed of light. 
If $v=c$, then \cite{Einstein05}:
\begin{equation}
\boxed{U = \frac{c+w}{1+w/c} = c.}
\end{equation}
Thus, the speed of light is invariant.

\subsection{Group Structure and Velocity Addition}

In his 1905 paper, Einstein writes that one can also obtain the formula for U \eqref{eq26} by composing two transformations of the form \eqref{eq12}. Introducing a third system $k'$ moving with velocity $w$ relative to $k$, the new transformation differs only in that it replaces $v$ with the equation \eqref{eq26}. He then remarks: “…one sees from this that such parallel transformations, as they must, form a group” \cite{Einstein05}

That is the extent of Einstein’s statement in his paper. He only claims closure of parallel (collinear) boosts. He points out that the formula for $U$ follows equally from composition, which shows that collinear boosts form a group. He does not spell out the intermediate transformations explicitly, i.e., no “from $k'$ to $k$ to $K$).
Here, I will reconstruct Einstein’s brief comment into a fuller group-theoretic derivation. We start with the Lorentz transformation from $K$ to $k$ \eqref{eq12}, where $k$ moves at velocity $v$ relative to $K$:
\begin{equation} \label{eq12-1}
\tau = \gamma_v\!\left(t - \frac{v}{c^{2}}x\right), 
\qquad
\xi = \gamma_v(x - vt),
\qquad 
\gamma_v = \frac{1}{\sqrt{1 - v^{2}/c^{2}}}.
\end{equation}
Now let $k'$ move at velocity $w$ relative to $k$. We apply the same transformation again: the transformation $k \to k'$ has the same form:
\begin{equation} \label{eq12-2}
\tau' = \gamma_w\!\left(\tau - \frac{w}{c^{2}}\xi\right), 
\qquad
\xi' = \gamma_w(\xi - w\tau),
\qquad
\gamma_w = \frac{1}{\sqrt{1 - w^{2}/c^{2}}}.
\end{equation}
We substitute the expressions for $\tau$ and $\xi$ in equation \eqref{eq12-1} into the transformation \eqref{eq12-2}:%
\footnote{This yields:
\begin{equation}
\begin{aligned}
\tau' 
&= \gamma_w\!\left(\tau-\frac{w}{c^{2}}\xi\right) \\
&= \gamma_w\!\left(\gamma_v\!\left(t-\frac{v}{c^{2}}x\right)
      -\frac{w}{c^{2}}\gamma_v(x-vt)\right) \\
&= \gamma_w\gamma_v\!\left[t-\frac{v}{c^{2}}x-\frac{w}{c^{2}}x
      +\frac{wv}{c^{2}}t\right] \\
&= \gamma_w\gamma_v\!\left[\left(1+\frac{vw}{c^{2}}\right)t
      -\frac{v+w}{c^{2}}\,x\right].
\end{aligned}
\end{equation}
\begin{equation}
\begin{aligned}
\xi' 
&= \gamma_w(\xi-w\tau) \\
&= \gamma_w\!\left(\gamma_v(x-vt)-w\,\gamma_v\!\left(t-\frac{v}{c^{2}}x\right)\right) \\
&= \gamma_w\gamma_v\!\left[x-vt-wt+\frac{wv}{c^{2}}x\right] \\
&= \gamma_w\gamma_v\!\left[\left(1+\frac{vw}{c^{2}}\right)x-(v+w)\,t\right].
\end{aligned}
\end{equation}}
This gives:
\begin{equation} \label{eq12-3}
\tau' = \gamma_w \gamma_v \left[ \left(1 + \frac{vw}{c^{2}}\right)t 
 - \frac{v+w}{c^{2}} x \right], \quad \xi' = \gamma_w \gamma_v \left[ \left(1 + \frac{vw}{c^{2}}\right)x 
 - (v+w) t \right].
\end{equation}
We factor the common prefactor:
\begin{equation} \label{eq12-4}
\Gamma_u = \gamma_v \gamma_w \left(1 + \frac{vw}{c^{2}}\right),
\end{equation}

\begin{equation} \label{eq12-5}
\tau' = \gamma_w \gamma_v \left(1 + \frac{vw}{c^{2}}\right) \left[t - \frac{v+w}{1+\frac{vw}{c^2}} \frac{x}{c^2} \right],
\end{equation}

\begin{equation} \label{eq12-6}
\xi' = \gamma_w \gamma_v \left(1 + \frac{vw}{c^{2}}\right) \left[x  - (v+w) t \right]=\gamma_v\gamma_w\!\left(1+\frac{vw}{c^{2}}\right)
\left[x-\frac{v+w}{\,1+\frac{vw}{c^{2}}\,}\,t\right].
\end{equation}
We now define:
\begin{equation} \label{eq41}
u = \frac{v+w}{1 + \tfrac{vw}{c^{2}}}.
\end{equation}
Then using equation \eqref{eq12-4} the transformations \eqref{eq12-5} and \eqref{eq12-6} take the form of \eqref{eq12} again:
\begin{equation} 
\tau' = \Gamma_u \left(t - \frac{u}{c^{2}}x\right),
\qquad
\xi' = \Gamma_u(x - ut).
\end{equation}
Finally, one verifies that the prefactor \eqref{eq12-4} is the usual Lorentz factor for $u$:
\begin{equation} \label{G}
\Gamma_u = \frac{1}{\sqrt{1 - u^{2}/c^{2}}},
\end{equation}
by inserting equation \eqref{eq41} and simplifying. Thus, the composed transformation has the same form as the original Lorentz transformation, but with $v$ replaced by $u$. 
Composing two collinear boosts of velocities $v$ and $w$ produces a boost of the same form with velocity \eqref{eq41}. 

Einstein remarks only that successive \emph{parallel} (collinear) transformations "wie dies sein mu\ss\ eine Gruppe bilden." He uses this as a kinematical check. He composes two collinear boosts yielding a transformation of the same form and reproduces the one-dimensional velocity-addition law.
Beyond this closure statement, he does not develop a general group-theoretic framework, i.e., identity $(v=0)$ and inverses $(v\mapsto -v)$ are implicit in the formulas, associativity is not discussed, and the paper does not treat the full Lorentz group (boosts with rotations) or its abstract structure.

\subsection{The Relation Without Interpretation}

In May 1905, Poincaré addressed Lorentz with courtesy, "Mon cher Collègue." He then noted, almost in passing: "Je trouve comme vous, $l=1$ par une autre voie." An agreement was reached, but not through \emph{ad hoc} dynamical assumptions; instead, it was achieved through the recognition of a group structure.
First, Poincaré introduced the Lorentz transformation in the form \cite{Wal}, letter 38.4:
\begin{equation} \label{eq97}
\boxed{x' = k l \, (x + \varepsilon t), \quad t' = k l \, (t + \varepsilon x), \quad y' = l y, \quad z' = l z,} 
\end{equation}
\begin{equation}
\text{with:} \qquad \boxed{k = \frac{1}{\sqrt{1 - \varepsilon^2}},}
\end{equation}
where $\varepsilon$ denotes a dimensionless velocity parameter, the velocity in units of the speed of light, set to $c=1$.
The Lorentz transformation gives us relations between coordinates $(x, t, y, z)$ and $(x', t', y', z')$.
Poincaré now composed two such transformations, first $(k,l,\varepsilon)$ and then $(k',l',\varepsilon')$.
The second map has the same form:
\begin{equation} \label{eq97-1}
x'' = k'l'(x' + \varepsilon' t'), \qquad
t'' = k'l'(t' + \varepsilon' x'), \qquad
y'' = l' y', \qquad
z'' = l' z'.
\end{equation}
He substitutes $x',t'$ from the first transformation \eqref{eq97} into the transformation \eqref{eq97-1}:

\begin{equation} \label{eq97-2} 
x'' = k'l'\!\left[\,k l(x+\varepsilon t) + \varepsilon' k l(t+\varepsilon x)\right]
= k k' l l' \left[ (1+\varepsilon\varepsilon') x + (\varepsilon+\varepsilon') t \right],    
\end{equation}

\begin{equation} \label{eq97-22}
t'' = k'l'\!\left[\,k l(t+\varepsilon x) + \varepsilon' k l(x+\varepsilon t)\right]
= k k' l l' \left[ (1+\varepsilon\varepsilon') t + (\varepsilon+\varepsilon') x \right].    
\end{equation}

\begin{equation}
y'' = l' y' = l l'\,y, \qquad z'' = l l'\,z .    
\end{equation}
He now demands that the result be again of the same form:
\begin{equation} \label{eq97-3}
x'' = k'' l'' (x+\varepsilon'' t), \qquad
t'' = k'' l'' (t+\varepsilon'' x), \qquad
y'' = l'' y, \qquad
z'' = l'' z ,
\end{equation}
for some parameters $k'',l'',\varepsilon''$.
From $y''=l''y=l'l\,y$ he reads off immediately \cite{Wal}, letter 38.4:
\begin{equation} \label{eq97-4}
\boxed{l'' = l\,l'.} 
\end{equation}
So the coefficient of $x$ in $x''$ must be proportional to $1$ and the coefficient of $t$ must be proportional to $\varepsilon''$. Similarly, in $t''$, the coefficient of $t$ must be proportional to $1$ and the coefficient of $x$ proportional to $\varepsilon''$.
So we can rewrite equations \eqref{eq97-2} and \eqref{eq97-22} as:

\begin{equation} \label{eq97-8}
\begin{aligned}
x'' &= k k' l l' (1+\varepsilon\varepsilon') 
\left[\, x + \frac{\varepsilon+\varepsilon'}{1+\varepsilon\varepsilon'}\, t \right], \\[1ex]
t'' &= k k' l l' (1+\varepsilon\varepsilon') 
\left[\, t + \frac{\varepsilon+\varepsilon'}{1+\varepsilon\varepsilon'}\, x \right].
\end{aligned}
\end{equation}
Thus, the comparison leads to \cite{Wal}, letter 38.4:
\begin{equation} \label{eq97-5}
\boxed{\varepsilon'' = \, \frac{\varepsilon+\varepsilon'}{1+\varepsilon\varepsilon'},} 
\qquad
k''l'' = kk'll'(1+\varepsilon\varepsilon').
\end{equation}
By definition, the Lorentz factor associated with parameter $\varepsilon''$ is \cite{Wal}, letter 38.4:
\begin{equation} \label{eq97-6}
\boxed{k'' = \frac{1}{\sqrt{1-\varepsilon''^{2}}}}.    
\end{equation}
We can now ask: does this $k''$ agree with the expression \eqref{eq97-3} implied by equations \eqref{eq97-2} and \eqref{eq97-22}? We plug the $\varepsilon''$ \eqref{eq97-5} into the $k''$ \eqref{eq97-6} and simplify:

\begin{equation}
k''=\frac{1+\varepsilon\varepsilon'}{\sqrt{(1-\varepsilon^{2})(1-\varepsilon'^{2})}} = \frac{1}{\sqrt{1-\varepsilon^2}} \, \frac{1}{\sqrt{1-\varepsilon'^2}}\,(1+\varepsilon\varepsilon')   
= k\,k'(1+\varepsilon\varepsilon'), 
\end{equation}
which matches exactly the expression implied by equations \eqref{eq97-8}, consistent with equation \eqref{eq97-4}.

Poincaré derived \eqref{eq97-4}. So, if each transformation has its own scale factor ($l, l'$), then the composition has $l''$ equal to the product of these two scale factors. Now, for the set of transformations to form a group with a single parameter $\varepsilon$, the scale factor must itself be a function of 
$\varepsilon$, say $l=f(\varepsilon)$. Then group closure requires:
\begin{equation} \label{eq97-9}
f(\varepsilon'')=f(\varepsilon)f(\varepsilon').    
\end{equation}
Poincaré suggested an ansatz, a possible form $f(\varepsilon)$ \cite{Wal}, letter 38.4:
\begin{equation} \label{eq97-10}
\boxed{l = (1 - \varepsilon^2)^m.}    
\end{equation}
If we insert that into equation \eqref{eq97-9}, we get: 
\begin{equation} \label{eq97-11}
(1 - \varepsilon''^2)^m \, = (1 - \varepsilon^2)^m \, (1 - \varepsilon'^2)^m.  
\end{equation}
But because of equation \eqref{eq97-5}, equation \eqref{eq97-11} becomes:%
\footnote{We start from equation \eqref{eq97-5}. Then:

\begin{equation}
1 - \varepsilon''^2=1 - \left(\frac{\varepsilon + \varepsilon'}{1 + \varepsilon \varepsilon'}\right)^2
= \frac{(1 + \varepsilon \varepsilon')^2}{(1 + \varepsilon \varepsilon')^2}
 - \frac{(\varepsilon + \varepsilon')^2}{(1 + \varepsilon \varepsilon')^2}
= \frac{(1 + \varepsilon \varepsilon')^2 - (\varepsilon + \varepsilon')^2}{(1 + \varepsilon \varepsilon')^2}.
\end{equation}
Expanding the numerator:
\begin{equation}
(1 + \varepsilon \varepsilon')^2 - (\varepsilon + \varepsilon')^2
= \left(1 + 2\varepsilon \varepsilon' + \varepsilon^2 \varepsilon'^2\right)
- \left(\varepsilon^2 + 2\varepsilon \varepsilon' + \varepsilon'^2\right)
= 1 - \varepsilon^2 - \varepsilon'^2 + \varepsilon^2 \varepsilon'^2 .
\end{equation}
Factoring:
\begin{equation}
1 - \varepsilon^2 - \varepsilon'^2 + \varepsilon^2 \varepsilon'^2
= (1 - \varepsilon^2)(1 - \varepsilon'^2). \qquad \text{Therefore:}
\end{equation}

\begin{equation} \label{eq97-12}
1 - \varepsilon''^2
= \frac{(1 - \varepsilon^2)(1 - \varepsilon'^2)}{(1 + \varepsilon \varepsilon')^2}. \quad \Rightarrow \quad (1 - \varepsilon''^2)^m = \frac{(1 - \varepsilon^2)^m(1 - \varepsilon'^2)^m}{(1+\varepsilon \varepsilon')^{2m}}.    
\end{equation}
Next, we substitute the right-hand side of equation \eqref{eq97-11} with $(1-\varepsilon''^2)^m  \, (1+\varepsilon \varepsilon')^{2m}$ from equation \eqref{eq97-12}.}

\begin{equation}
(1 - \varepsilon''^2)^m \, = (1 - \varepsilon''^2)^m \, (1+\varepsilon \varepsilon')^{2m}.   
\end{equation}
For this to hold for all $\varepsilon, \varepsilon'$, we must have $m=0$ \cite{Wal}, letter 38.4. Hence:
\begin{equation}
(1 - \varepsilon''^2)^m=1.    
\end{equation}

In his paper "On the Dynamics of the Electron," Poincaré repeated this derivation but then took the decisive extra step. He concluded that the only consistent choice is $l=1, l'=1, l''=1$. Thus, the only possibility compatible with the group property is $l=1$. Therefore, in the "Dynamics of the Electron" memoir, Poincaré does what he did not yet do in the letter: he closes the argument, showing that the group property itself rules out any $l\neq 1$ \cite{Poi05-2}.

\subsection{The Formula Versus the Law}

Miller observes that “considering two successive Lorentz transformations along the same direction, it was easy for [Poincaré] to prove that the Lorentz transformations form a group, and must be equal to one. As a bonus, Poincaré also obtained the new addition law for velocities that is independent of $l$” \cite{Miller-1}. In his recent book, \emph{Poincaré, Einstein and the Discovery of Special Relativity. An End to the Controversy}, Ginoux echoes this claim, stating that “Einstein thus obtains the new relativistic velocity addition law established a few weeks earlier by Poincaré” (in section \S 5) \cite{Gin}. Likewise, Olivier Darrigol, in \emph{Relativity Principles and Theories from Galileo to Einstein}, asserts that Poincaré “shows that the product of two parallel boosts of velocities $u$ and $v$ is a boost of velocity…,” and he reproduces Poincaré’s equation \eqref{eq97-5} in Einstein’s notation for the velocity-addition law \eqref{eq41} \cite{Dar}.
However, the resemblance of form should not be mistaken for an identity of meaning or function.
Equation \eqref{eq97-5} is the same linear form as Einstein’s one-dimensional addition law for velocities \eqref{eq41}, once we identify $\varepsilon=\frac{v}{c}$. Yet the role it plays in each context is quite different.  

In his 1905 paper, Einstein observes that the velocity $U$ given in equation \eqref{eq26} can also be obtained by composing two transformations of the form~\eqref{eq12}. He considers a third system $k'$ [equation \eqref{eq12-2}] moving with velocity $w$ relative to $k$ and notes that the resulting transformation differs only in that it replaces $v$ with the expression for $U$. He then remarks that “one sees from this that such parallel transformations, as they must, form a group” \cite{Einstein05}. This is the entirety of Einstein’s comment: he claims closure of collinear boosts, pointing out that the formula for $U$ follows equally from composition, which shows that successive transformations preserve their form.  

Suppose one reconstructs Einstein’s brief statement into a complete derivation. In that case, the procedure is straightforward: take the Lorentz transformation from $K$ to $k$ with parameter $v$, insert it into the transformation from $k$ to $k'$ with parameter $w$, and simplify. The result is a transformation of the same form with parameter equation \eqref{eq41}, and the prefactor is exactly the Lorentz factor \eqref{G} associated with $u$. Thus, the composition of two collinear boosts of velocities $v$ and $w$ is another boost with velocity $u$, reproducing the one-dimensional velocity–addition law \eqref{eq26}.  

It is essential, however, to note what Einstein does and does not do. He presents the calculation only as a \emph{kinematical check}, remarking on the group property of parallel transformations but leaving the general group structure undeveloped: the identity transformation $(v=0)$ and inverses $(v\mapsto -v)$ remain implicit, associativity is not discussed, and the treatment is restricted to collinear boosts. Poincaré, by contrast, had earlier written the Lorentz transformations with a dimensionless parameter $\varepsilon=v/c$ and showed that their composition produces a new transformation with parameter equation \eqref{eq97-5}, remarking that the transformations “form a group.” But whereas Poincaré presented this as a \emph{structural property} of the Lorentz transformations, treating $\varepsilon$ as an abstract group parameter, Einstein explicitly interpreted the formula as the relativistic law of velocity addition, a relation he had already derived independently by kinematical means.

\medskip
What has been established here is the first part of the research. Part two awaits in the sequel.

\appendix
\section{Appendix: Alternative Derivation of the Relativity of Simultaneity} \label{AP}

The derivation builds up the isotropy ---  two events are simultaneous if the light signal emitted from $A$ and reflected at $B$ takes equal time to travel $A \rightarrow B$ and $B \rightarrow A$ --- explicitly in coordinates. 
It is mathematically equivalent and conceptually aligned with Einstein's derivation [equations \eqref{A}, \eqref{B}, \eqref{C}, \eqref{D}, \eqref{eq115}, \eqref{A-1}, \eqref{B-1}, \eqref{C-1}, \eqref{D-1}, and \eqref{eq116}]. This derivation shares the same algebraic style as his operational definition of simultaneity in section \S 1 and his derivation of the Lorentz transformation in \S 3. Here is the derivation:

Let us first calculate the outbound trip $A \to B$. At time $t=t_A$, a light pulse is emitted from point $A$, so at the instant of emission, the light’s position is: 

\begin{equation} \label{eq117}
x_A(t_A).    
\end{equation}
From that instant onward, the light propagates in the $+x$-direction at speed $V$. So, after a time interval $t-t_A$, it has traveled a distance:
\begin{equation} \label{eq118}
V(t-t_A).    
\end{equation}
Adding equation \eqref{eq118} to the starting position \eqref{eq117} gives the position of the light after emission:
\begin{equation} \label{eq119}
x_\ell(t) = x_A(t_A) + V(t-t_A).
\end{equation}
Now, the initial position of $B$ at emission of light is: 
\begin{equation} \label{eq120}
x_B(t_A) =  x_A(t_A) + r_{AB}.  
\end{equation}
$B$ is in uniform motion with velocity $v$ for $t>t_A$, thus:
\begin{equation} \label{eq121}
x_B(t) =  x_B(t_A) +  v(t-t_A).
\end{equation}
Substituting equation \eqref{eq120} into \eqref{eq121} gives:
\begin{equation} \label{eq122}
x_B(t_B) =  x_A(t_A) + r_{AB} +  v\, (t_B-t_A).
\end{equation}
This is the position of $B$ after emission.

\noindent At the specific moment when the light reaches $B$, $t=t_B$. Thus:
\begin{equation} \label{eq119-1}
x_\ell(t_B) = x_A(t_A) + V(t_B-t_A).
\end{equation}
This equation is just the special case of equation \eqref{eq119}, evaluated at $t=t_B$. Therefore, we can equate it to the position of $B$ at that instant: $x_B(t_B)=x_\ell(t_B)$, i.e., equation \eqref{eq119-1} $=$ equation \eqref{eq122}:
\begin{equation}
x_A(t_A) + V\,(t_B-t_A) = x_A(t_A) + r_{AB} + v\,(t_B-t_A).
\end{equation}
We cancel $x_A(t_A)$ and rearrange:%
\footnote{Subtract $v\,(t_B - t_A)$ from both sides $V\,(t_B - t_A) - v\,(t_B - t_A) = r_{AB}$, and factor out $(t_B - t_A)$ on the left.}
\begin{equation}
V\,(t_B-t_A) = r_{AB} + v\,(t_B-t_A)
\;\;\Longrightarrow\;\; (V-v)(t_B-t_A) = r_{AB}. 
\end{equation}
Hence, from the perspective of system $K$, the travel times of a light signal between $A$ and $B$ are represented by equation \eqref{eq115}.

Now we calculate the return trip $B \to A$. After reflection of light at $t_B$, the light heads back with:
\begin{equation} \label{eq123}
x_\ell(t) = x_B(t_B) - V\,(t-t_B).
\end{equation}
Meanwhile, $A$ keeps moving:
\begin{equation} \label{eq124}
x_A(t) = x_A(t_A) + v\,(t-t_A).
\end{equation}
We can equate equation \eqref{eq123} and \eqref{eq124} to obtain
at the reception of the light signal $t=t'_A$:
\begin{equation} \label{eq125}
x_B(t_B) - V\,(t'_A-t_B) = x_A(t_A) + v\,(t'_A-t_A). 
\end{equation}
We then substitute $x_B(t_B)$ [equation \eqref{eq122}] into equation \eqref{eq125}, and cancel $x_A(t_A)$:
\begin{equation}
r_{AB} + v\,(t_B-t_A) - V\,(t'_A-t_B) = v\,(t'_A-t_A).
\end{equation}
Bring the $v$-terms together:
\begin{equation}
r_{AB} - V\,(t'_A-t_B) = v\Big[(t'_A-t_A)-(t_B-t_A)\Big] = v\,(t'_A-t_B). 
\end{equation}
This finally yields \eqref{eq116}.

Einstein’s definition of synchronization requires equation \eqref{eq114}, which implies equation \eqref{eq1}:
\begin{equation} \label{eq129}
t_B = \tfrac12\,(t_A+t'_A) \;\equiv\; t_m.
\end{equation}
Thus, in a synchronized system, the reflection at $B$ occurs at the midpoint time $t_m$. For the system $k$, however, one finds $t_B \neq t_m$, so that the clocks do not remain synchronous.

Einstein’s demonstration of the relativity of simultaneity relies on equations \eqref{eq115} and \eqref{eq116}. If we instead apply the Lorentz transformation directly, the relativity of simultaneity appears in a single line. From the time transformation \eqref{eq2} one finds:
\begin{equation} \label{eq4}
\Delta t' = \gamma \!\left(\Delta t - \frac{v}{c^2}\,\Delta x\right).
\end{equation}
For two distinct events that are simultaneous in $K$ ($\Delta t=0$, $\Delta x\neq 0$), equation \eqref{eq4} reduces to:
\begin{equation} \label{eq3}
\Delta t' = -\,\gamma\,\frac{v}{c^2}\,\Delta x \;\neq\; 0.
\end{equation}
Let us connect this with Einstein’s argument. 
\noindent From equations \eqref{eq115} and \eqref{eq116} we can first subtract:
\begin{equation}
(t_B - t_A) - (t'_A - t_B)
= \frac{r_{AB}}{V - v} - \frac{r_{AB}}{V + v}
= r_{AB}\!\left(\frac{1}{V - v} - \frac{1}{V + v}\right).
\end{equation}
Then we combine the fractions:
\begin{equation}
\frac{1}{V - v} - \frac{1}{V + v}
= \frac{(V+v) - (V-v)}{(V-v)(V+v)}
= \frac{2v}{V^2 - v^2}.
\end{equation}
Thus, the asymmetry of the path times is:
\begin{equation} \label{eq132}
(t_B-t_A)-(t'_A-t_B)=\frac{2v\,r_{AB}}{V^2-v^2}.
\end{equation}
Taking into consideration equation \eqref{eq1} or \eqref{eq129}, the $K$–time gap $\Delta t_K$ between the reflection at $B$ and the midpoint at $A$ is:
\begin{equation} \label{eq133}
\Delta t_K \equiv t_B-t_m 
= \tfrac12\!\left[(t_B-t_A)-(t'_A-t_B)\right].
\end{equation}
Substituting equation \eqref{eq132} into \eqref{eq133} yields: 
\begin{equation} \label{eq134}
\Delta t_K = \frac{v}{V^2-v^2}\,r_{AB}. 
\end{equation}
Transforming equation \eqref{eq129} to system $k$ (using equation \eqref{eq134}) gives:
\begin{equation}
\Delta t'=\gamma\!\left(\Delta t_K-\frac{v}{V^2}r_{AB}\right)
=\gamma\!\left(\frac{v}{V^2-v^2}\,r_{AB}-\frac{v}{V^2}\,r_{AB}\right). 
\end{equation}
Simplifying, we get:
\begin{equation}
\Delta t'=-\,\gamma\,\frac{v}{V^2}\,\frac{V^2}{V^2-v^2}\,r_{AB}
=-\,\frac{v}{V^2}\,\gamma^2\,r_{AB}.
\end{equation}
If we now identify the proper length of the rod as $L_0 \equiv \Delta x=\gamma\,r_{AB}$, this becomes equation \eqref{eq3}. Thus, Einstein’s equations \eqref{eq115} and \eqref{eq116} and the Lorentz transformation result are mathematically equivalent, although he presents only the former in the 1905 paper.

\end{document}